\documentclass[11pt]{article}

\usepackage{epsfig}
\usepackage{amsmath}
\usepackage{amssymb}

\begin{document}


\renewcommand{\thefootnote}{\fnsymbol{footnote}}

\begin{center}
\textbf{\large Phase diagram of QCD with four quark flavors
at finite temperature and baryon density} \\
\vspace{0.75truecm}
Vicente Azcoiti$^a$\footnote{E-mail: azcoiti@azcoiti.unizar.es}, 
Giuseppe Di Carlo$^b$\footnote{E-mail: dicarlo@lngs.infn.it}, 
Angelo Galante$^{b,c}$\footnote{E-mail: galante@lngs.infn.it}, 
Victor Laliena$^a$\footnote{E-mail: laliena@unizar.es} \\
\vspace{0.5truecm}
{\footnotesize
\textit{$^a$ Departamento de F\'{\i}sica Te\'orica, Universidad de
Zaragoza, \\
Pedro Cerbuna 12, E-50009 Zaragoza (Spain)}
\\
\vspace{0.15truecm}
\textit{$^b$ INFN, Laboratori Nazionali del Gran Sasso, 
67010 Assergi (L'Aquila) (Italy)} \\
\vspace{0.15truecm}
\textit{$^c$ Dipartimento di Fisica dell'Universit\`a di L'Aquila
67100 L'Aquila (Italy)}
}
\\
\vspace{0.5truecm}
April 27, 2005
\end{center}

\vspace{1truecm}

\begin{center}
\textbf{Abstract}
\end{center}

\noindent
We analyze the phase diagram of QCD with four staggered 
flavors in the $(\mu, T)$ plane using a method recently proposed by us.
We explore the region $T\gtrsim 0.7 \, T_\mathrm{C}$ and 
$\mu\lesssim 1.4 \, T_\mathrm{C}$, 
where $T_\mathrm{C}$ is the transition temperature at zero baryon density,
and find a first order transition line.
Our results are quantitatively compatible with those obtained 
with the imaginary chemical potential approach
and the double reweighting method, 
in the region where these approaches are reliable, 
$T\gtrsim 0.9 T_\mathrm{C}$ and $\mu\lesssim T_\mathrm{C}$. 
But, in addition, our method allows us to extend the transition line to lower 
temperatures and higher chemical potentials.

\vspace{0.5truecm}


\vfill\eject


\renewcommand{\thefootnote}{\arabic{footnote}}
\setcounter{footnote}{0}

\section*{\normalsize 1 \ Introduction}

The study of QCD at finite baryon density is one elusive problem of
utmost importance for the understanding of strong interactions.
Many features of the behavior of hadronic and quark matter at high
baryon density have been suggested mainly from the analysis of 
effective theories. 
On this basis, a rich structure of phases and phase transitions is 
expected, and the dependence of the phase structure on the quark 
masses has been conjectured \cite{kogut:book}.
Unfortunately, the sign problem prevents the direct use of Monte Carlo 
simulations at finite baryon density. However, in the last years
methods to get the transition line at high
temperature and low baryon density, overcoming the sign problem,
have been proposed.
One method uses double reweighting (in the two parameters, 
$\beta$ and $\mu a$) of the configurations
generated at the transition temperature at zero chemical potential,
following the transition line in the $(\mu a,\beta)$ plane~\cite{fodor}. 
The rational of this method is that it is expected that the overlap 
between the reweighted ensemble and the ensemble at a given 
point of the transition line will be improved respect to the 
Glasgow method~\cite{glasgow}, since the original ensemble is itself
a mixture of configurations corresponding to the confined and deconfined
phase. This approach has been used in a first attempt to locate the
expected critical endpoint of QCD with 2+1 flavors by using only
first principles~\cite{fodor2p1}.
Another method exploits the well known fact that there is no 
sign problem if the chemical potential is purely imaginary~\cite{imaginary}.
Then, it was realized in~\cite{philippe} that it is possible to determine, 
by means of numerical simulations,
a pseudo-transition line at imaginary chemical potential and 
extend it analytically to real chemical potential. 
This method has been used to study the phase diagram at small chemical
potential around the zero density transition point in QCD with two 
and three degenerate quark flavors in~\cite{philippe} and with four 
flavors in~\cite{mp}.
Another proposal that is being employed to extract information about the
phase transition in the region of small chemical potential 
is to compute the expectation values of the
derivatives respect to $\mu a$ at $\mu a=0$, in order to reconstruct
several terms of the Taylor series~\cite{taylor,gavai}.

Recently 
we devised another method which can be regarded as a generalization of 
the imaginary chemical potential approach. The former has several 
advantages over the later,
for it seems that it can be used to determine the transition line at lower
temperatures and higher densities~\cite{JHEP}. Especially interesting is the
fact that it may be used to locate
the critical endpoint expected in two flavor QCD~\cite{crit}.

In this paper we report the results of the phase transition line of lattice
QCD with four degenerate flavors of staggered quarks obtained with this 
new method. At zero density a very clear first order transition 
separates the low
and high temperature phases, at a transition temperature, $T_\mathrm{C}$, 
that, for 
small quark masses, ranges from 100~MeV to 170~MeV \cite{Tzeromu}. 
It is expected that
the transition continues along a line in the $(\mu,T)$ plane, with the
transition temperature lowering as $\mu$ increases. This is what has been
obtained using the double reweighting~\cite{fodor} and the imaginary
chemical potential~\cite{mp} approaches. 
We also find a first order transition line starting at $\mu=0$ and 
$T=T_\mathrm{C}$
and continuing at lower temperatures as $\mu$ increases.
Our results are quantitatively compatible with those obtained in~\cite{mp}
with the imaginary chemical potential approach
and with the results of the double reweighting method \cite{fodor}, 
in the region where these approaches are reliable 
(see sections three and four for a discussion on this point).
But, in addition, our method allows us to extend the transition line to lower 
temperatures and higher chemical potentials.

The remaining of the paper is organized as follows: in next section we 
explain our approach. In section three we describe the numerical
results and in section four we present our conclusions.

\section*{\normalsize 2 \ Review of the method}

The numerical method used in this work is based on the definition of a 
generalized QCD action which depends on two free parameters $(x, y)$. This 
generalized action suffers also from the sign problem for real values of $y$ 
but not for imaginary values of it. Simulations will be then 
performed at imaginary values of $y$ and at the end analytical extensions will 
be needed. The main advantage of this approach when compared with the 
imaginary chemical potential method is that we can explore the  
phase transition line at imaginary values of $y$ at any given physical 
temperature i.e., we are not forced, as in the case of imaginary chemical 
potential, to perform simulations at so high temperatures that the system is 
in the quark-gluon plasma phase for any real value of $\mu a$.
In this section we shall describe this method. 
The interested reader can find more details of it in~\cite{JHEP}. 

The lattice action for QCD with staggered fermions and 
chemical potential $\mu$ is

\begin{eqnarray}
S &=& S_\mathrm{PG} + \frac{1}{2}\sum_n\sum^3_{i=1}
\bar\psi_n \eta_i (n) \left( U_{n,i}
\psi_{n+i} - U^\dagger_{n-i,i}\psi_{n-i}\right) \nonumber \\
&+& \frac{1}{2}\sum_{n} \bar\psi_n \eta_0 (n) 
\left( e^{\mu a}  U_{n,0}\psi_{n+0}
- e^{-\mu a} U^\dagger_{n-0,0}\psi_{n-0}\right) 
+ ma\sum_n \bar\psi_n\psi_n\, ,
\label{action}
\end{eqnarray}

\noindent
where $S_\mathrm{PG}$ is the standard Wilson action for the gluonic fields,
which contains $\beta$, the inverse gauge coupling, as a parameter,
$\eta_i (n)$ and $\eta_0(n)$ are the Kogut-Susskind phases,
$m$ the fermion mass, and $a$ the lattice spacing.

Let us now define the following generalized action

\begin{equation}
S = S_\mathrm{PG} + ma \sum_n \bar\psi_n\psi_n
+ \frac{1}{2}\sum_n\sum^3_{i=1}\bar\psi_n \eta_i (n)
\left( U_{n,i} \psi_{n+i} - U^\dagger_{n-i,i}\psi_{n-i}\right) +
S_\tau(x, y)\, ,
\label{gaction}
\end{equation}

\noindent
with

\begin{eqnarray}
S_{\tau}(x, y) &=& x \frac{1}{2} \sum_{n}\bar\psi_n \eta_0 (n)\left( U_{n,0}
\psi_{n+0} - U^\dagger_{n-0,0}\psi_{n-0}\right) \nonumber \\
&+& y \frac{1}{2} \sum_{n} \bar\psi_n \eta_0 (n)\left(  U_{n,0}
\psi_{n+0} + U^\dagger_{n-0,0}\psi_{n-0}\right)\, ,
\label{actiontxy}
\end{eqnarray}

\noindent
where $x$ and $y$ are two independent parameters. The QCD action is recovered
by setting $x= \cosh(\mu a)$ and $y= \sinh(\mu a)$.

\begin{figure}[t!]
\centerline{\includegraphics*[width=2.5in,angle=270]{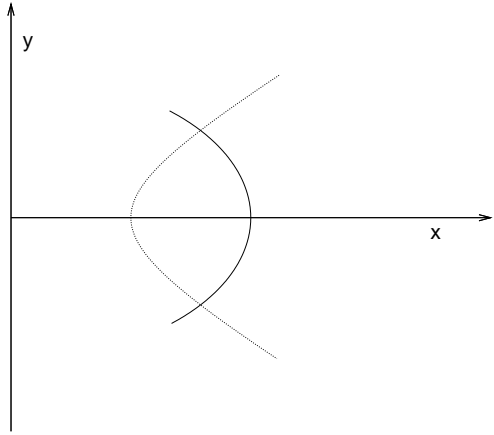}}
\caption{Minimal phase diagram conjectured for the generalized QCD.
The solid line is line of phase transitions. The discontinuous line is the
physical line, $x^2-y^2=1$.}
\label{fig:phd_xy}
\end{figure} 

Monte Carlo simulations of the model (\ref{gaction}) for real values of
$x, y$ are not feasible since we meet the sign
problem. However if $y$ is a pure imaginary number,
$y= i\bar y$, where $\bar y$ is real, the sign problem disappears
since the fermionic matrix is the
sum of a constant diagonal matrix plus an antihermitian matrix which
anticommutes with the staggered version of the $\gamma_5$ Dirac matrix, 
and numerical simulations become feasible.
Imaginary chemical potential is a particular case of this, obtained by
setting $x=\cos(\mu a)$ and $\bar{y}=\sin(\mu a)$. 

The expected phase diagram for this model in the $(x, y)$ plane  
is shown in Fig.~\ref{fig:phd_xy} \cite{JHEP}. 
The solid line is a presumed line of phase
transitions and the discontinuous
line is the physical line, $x^2 - y^2 = 1$, along
which one recovers standard QCD at finite baryon density. 
The intersection of the solid line with the discontinuous one will
therefore give us
the transition chemical potential of QCD at a given temperature. A
change in the physical temperature can be simulated by changing $\beta$
keeping fixed $L_t$ or vice-versa. In both cases the solid line in 
Fig.~\ref{fig:phd_xy}
will move and the intersection point which gives the transition chemical
potential will change with the physical temperature.
For small $\beta$ the transition line crosses the $y=0$ axis at 
$x>1$ and, therefore, intersects the physical line also at
$x>1$, producing a physical phase transition at $\mu a >0$. 
By increasing $\beta$ and keeping fixed the
temporal lattice extent $L_t$, the transition point on the
$y= 0$ axis moves toward $x = 1$ and eventually crosses it. 
Clearly, the value of $\beta$ at which the transition line intersects
the physical line at $x=1$ and $y=0$ is the zero density transition point.
For larger $\beta$ the transition line and the physical line do not intersect
and the system is in the deconfined phase whatever the chemical potential.

>From an analysis of the symmetries of the action~(\ref{gaction})
it is not difficult to realize that the partition function depends 
on $x$ and $y$ only through the combinations

\begin{eqnarray}
u &=& x^2-y^2\, , \nonumber \\
v &=& (x+y)^{3L_t}\:+\:(x-y)^{3L_t}\, .
\end{eqnarray}

\noindent 
For imaginary values of $y$ ($y=i\bar{y}$) we have $u=\rho^2$
and $v=2\rho^{3L_t}\cos(3L_t\eta)$, where $\rho^2=x^2+\bar{y}^2$
and $\tan\eta=\bar{y}/x$, and, therefore, the free energy will be a 
periodic function of $\eta$ with period $2\pi/3L_t$. In
particular if the phase transition line of Fig. 1 continues to imaginary
values of $y$, the expected phase diagram in the $(x, \bar{y})$ plane
is displayed in Fig. 2, where we have incorporated the property of
periodicity.
In Fig. 2 we have also included the line $\rho = 1$ (dotted) which
is the locus of the points accessible to numerical simulations of QCD at
imaginary chemical potential. One can see
now how this approach has more potentialities than the
imaginary chemical potential approach. Indeed by increasing the inverse
gauge coupling $\beta$, the phase transition line of Fig. 2 moves toward
the origin of coordinates. In some interval $(\beta_m, \beta_M)$
the transition line intersects the $\rho=1$ line and then a phase transition
will appear at imaginary chemical potential. In such a situation, the physical
temperature is so high that the system is in an unconfined phase for any real
value of the chemical potential. 
The advantage of our approach is that in our simulations $\rho$ is not 
enforced to be one.

The variables $u$ and $v$ have the interesting property that are real
for both $y$ real and pure imaginary, so that they map the two planes 
$(x,y)$ and
$(x,\bar{y})$ onto a single plane $(u,v)$. The line $v=2u^{3L_t/2}$
separates the regions corresponding to each plane: the region above it
in Fig.~3 corresponds to real $y$ and is not accessible to
numerical simulations, while the region below it corresponds to imaginary
$y$ and can be explored by means of simulations. The physical line
is $u=1$ with $v\geq 2$. The imaginary chemical potential points are
mapped onto the line $u=1$, $-2\leq v\leq 2$. The analytical extension
from imaginary $y$ to the physical real $y$ becomes in the $(u,v)$
plane the extrapolation from the region accessible to numerical simulations to
$u=1$, $v\geq 2$.

\begin{figure}[t!]
\centerline{\includegraphics*[width=2.5in,angle=270]{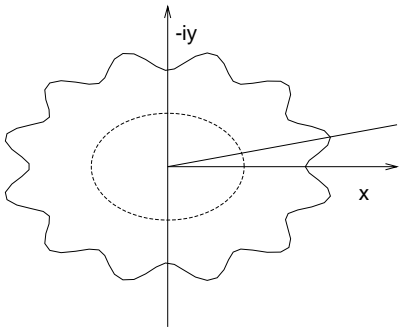}}
\vspace{-0.5truecm}
\caption{Conjectured phase diagram in the $(x,\bar{y})$ plane.
We have incorporated the periodicity. The dashed line contains the
points corresponding to imaginary chemical potential.}
\label{fig:phd_xyimag}
\end{figure} 

In order to get the transition chemical potential we must
determine the coordinates of the intersection point between the solid line
and the physical line in Fig.~\ref{fig:phd_xy}. 
>From the symmetries of the partition
function we know that the phase transition line is an
even function of $x$ and $y$. Therefore, we
can write the following equation for the transition 
line at fixed $\beta$ in the $(x,y)$ plane\footnote{Here and in the
following we do not write explicitly the dependence of the coefficients
$a_i$,$\beta_0$,$b_i$, etc. on the temporal lattice extent, $L_t$, and
the quark masses, $ma$.}

\begin{equation}
x^2 = 1 + a_0(\beta) + a_2\left(\beta\right) y^2
+ a_4\left(\beta\right) y^4
+ O(y^6),
\label{criticalline}
\end{equation}

By fixing the lattice temporal extent $L_t$ and the gauge coupling $\beta$
one fixes the physical temperature $T$. The intersection point of the
transition line with the physical line

\begin{equation}
y^2_c = a_0\left(\beta\right) + a_2\left(\beta\right) y^2_c
+ a_4\left(\beta\right) y^4_c
+ O(y^6_c),
\label{criticalmu}
\end{equation}

\noindent
will give us the transition chemical potential, $y_c=\sinh(\mu_c a)$, at this
temperature.

The strategy for the determination of the transition chemical potential is
then the following. From numerical simulations at imaginary values of $y$,
$y=i\bar y$, near the phase transition point $((1+a_0)^{1/2}, 0)$
one can locate several
phase transition points in the $(x, \bar y)$ plane (see Fig. 2). By fitting
these points with equation (\ref{criticalline}) with the $+$ sign of the
coefficient proportional to $y^2$ replaced by $-$, we can numerically measure
the first coefficients. Ignoring the quartic term, 
the transition chemical potential, $\mu_c a$, will then be given by

\begin{equation}
\mu_c a = \pm \sinh^{-1} \left(\frac{a_0}{1-a_2}\right)^{1/2}.
\label{muc}
\end{equation}

An alternative procedure, which is indeed the one employed in this work,  
is to project of the phase diagram onto 
the $(y,\beta)$ plane. In practice, we fix $x=x_0>1$, $L_t$ and the lattice
quark masses\footnote{Notice that in the imaginary chemical potential approach
$x_0$ is enforced to be less than one.} and perform simulations for different
values of $\bar{y}$ and $\beta$.
In this way we can easily find an accurate estimate of the transition point
at a given $\bar{y}$ by interpolation in $\beta$ via reweighting.
The qualitative phase diagram in the $(y,\beta)$ plane is
displayed in Fig.~4. The solid lines are the physical lines,

\begin{equation}
y = y_\mathrm{ph} = \pm\sqrt{x_0^2-1}\, ,
\end{equation}

\noindent
and the discontinuous line is a line of phase transitions.

Using again the symmetries
of the partition function we can write for this line the following 
equation

\begin{equation}
\beta = \beta_0\left(x_0\right)  + b_2 \left(x_0\right) y^2
+ O\left(y^4\right)\, .
\label{clineybeta0}
\end{equation}

As in the case discussed previously, one can measure $\beta_0$ and $b_2$ from
simulations at $y=0$ and at imaginary $y=i\bar y$ (keeping $x_0$ fixed).
Then, we have to find the intersection point between the
physical and phase transition lines of Fig.~4.

Equivalently, one may use the plane $(u,\beta)$, where $u=x_0^2+\bar{y}^2>1$.
>From simulations at fixed $x=x_0$ and $\bar{y}\geq 0$ we can get a phase 
transition line $\beta(u)$ for $u\geq x_0^2 > 1$. 
Then one can extrapolate this line to $u=1$, thus obtaining the physical 
transition coupling at chemical potential $\mu a=\cosh^{-1}(x_0)$.
 

\begin{figure}[t!]
\centerline{\includegraphics*[width=2.0in,angle=270]{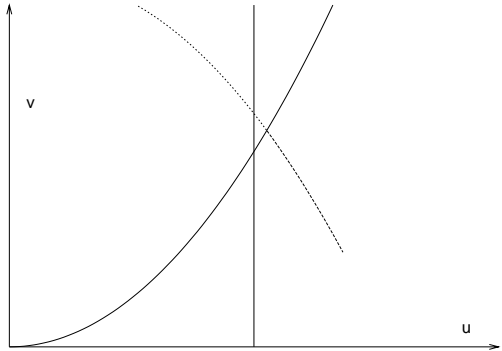}}
\caption{Phase diagram in the $(u,v)$ plane. The solid line,
$v=2u^{3L_t/2}$ corresponds to $y=0$, and separates the region
where numerical simulations are feasible (below the line) from the 
region where the sign problem prevents numerical simulations
(above the line). The discontinuous line is a hypothetical phase
transition line. The line $u=1$ is also displayed. For $v\geq 2$
it is the physical line while for $-2\leq v\leq 2$ it corresponds
to imaginary chemical potential.}
\label{fig:phd_uv}
\end{figure}

\section*{\normalsize 3 \ Numerical results}

We performed simulations of QCD with four degenerate flavors of staggered
quarks using the Hybrid Monte Carlo algorithm on lattices of sizes
$6^3\times 4$ and $8^3\times 4$. Our method is computationally
very expensive and it is rather difficult to go to larger lattices.
The quark mass in lattice units was fixed to $m a=0.05$.
On each lattice, we repeated the simulations for several values of $x$, 
$\bar{y}$,
and $\beta$. We measured the plaquette, the chiral condensate, and the
Polyakov loop after each molecular dynamics trajectory of unit time.
For each simulation we accumulate between thirty and forty thousand
measurements. Most of the simulations have been performed on the 
Linux clusters of LNGS-INFN, and some of them in the Linux cluster
of Departamento de F\'{\i}sica Te\'orica of Universidad de Zaragoza.
 
\begin{figure}[t!]
\centerline{\includegraphics*[width=2.0in,angle=90]{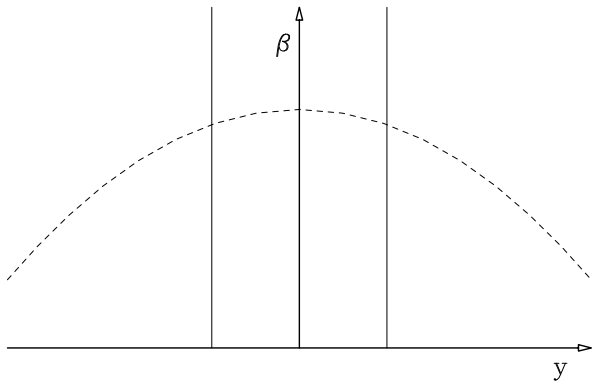}}
\caption{Conjectured phase diagram in the $(y,\beta)$ plane,
with $x=x_0>1$ fixed.
The discontinuous line is a line of phase transitions, whereas
the solid lines are the physical lines, 
$y=y_\mathrm{ph}=\pm\sqrt{x_0^2-1}$.}
\label{fig:phd_betay}
\end{figure} 

At zero chemical potential ($x=1$ and $\bar{y}=0$)
there is a very clear signal of a first order
phase transition controlled by $\beta$, at which the plaquette, the chiral
condensate, and the Polyakov loop vary abruptly and show a clear two state
structure. This first order transition persists for $x>1$ and $\bar{y}\geq0$,
with the transition coupling depending on $x$ and $\bar{y}$.
To determine the transition lines numerically, we proceeded as follows.
For each pair $(x,\bar{y})$ we estimate the transition coupling
by reweighting \textit{\`a la} Ferrenberg-Swendsen (in the parameter $\beta$)
the configurations corresponding to the value of $\beta$ at which 
a clear two state signal appeared. We used the maximum of the plaquette 
susceptibility as the criterion to define the transition coupling on the
finite lattice, since it gives the best signal.

In this way, for each value of $x$ we have a transition line in the plane
$(\bar{y},\beta)$. We fit this line with a second order polynomial:
\begin{equation}
\beta(\bar{y})\;=\;\beta_0(x)\:+\:b_2(x)\bar{y}^2\, .
\label{imag_line}
\end{equation}
The analytic continuation of this functions to real $\bar{y}=-iy$ is trivial:
\begin{equation}
\beta(y)\;=\;\beta_0(x)\:-\:b_2(x) y^2\, .
\end{equation}
The physical value of $y$ is $y_{\mathrm{ph}}=\sqrt{x^2-1}$.
Hence, the physical transition coupling is  
$\beta(y_{\mathrm{ph}})=\beta_0(x)-b_2(x) (x^2-1)$.

We can also use the variable $u=x^2+\bar{y}^2$, and then the transition
line in the plane $(u,\beta)$ is fit by the linear function
\begin{equation}
\beta(u)\;=\;\beta_\mathrm{ph}(x)\:+\:b_2(x)\,(u-1)\, .
\label{uline}
\end{equation}

Notice that the fits in the $(\bar{y},\beta)$ and $(u,\beta)$ are
identical since both the data and the fit functions are related
by the same transformation.
In the $(u,\beta)$ plane the analytic continuation to 
$\bar{y}=-\mathrm{i}y_\mathrm{ph}$ corresponds to the extrapolation to $u=1$.
Hence, the physical transition coupling for each value
of $x$ is directly given by the parameter $\beta_\mathrm{ph}$.

\begin{table}[t!]
\centerline{
\begin{tabular}{llllllc}
\multicolumn{1}{c}{$L$} &
\multicolumn{1}{c}{$x$} & \multicolumn{1}{c}{$\beta_0$} & 
\multicolumn{1}{c}{$\beta_{\mathrm{ph}}$} &
\multicolumn{1}{c}{$b_2$} & \multicolumn{1}{c}{$\chi^2/\mathrm{ndof}$}
& \multicolumn{1}{c}{ndof}  \\
\hline\hline 
6 & 1.02  & 5.0359(2) & 5.008(1) & 0.68(2) & \ \ 0.4  & 1 \\
6 & 1.045 & 5.028(1)  & 4.962(3) & 0.72(3) & \ \ 1.63 & 1 \\
\hline
8 & 1.0201 \ \  & 5.0367(5) \ \  & 5.0089(10) \ \  & 0.684(15) \ \  
& \ \ 0.82  & 4 \\
8 & 1.0314 & 5.0332(3) & 4.9869(12) & 0.726(13)  & \ \ 1.11 \ \ & 5 \\
8 & 1.0453 & 5.0292(5) & 4.9636(21) & 0.708(21)  & \ \ 0.97 \ \ & 5 \\
8 & 1.0811 & 5.0179(4) & 4.8943(20) & 0.732(10)  & \ \ 0.36 \ \ & 7 \\
8 & 1.1276 & 5.0047(4) & 4.8017(28) & 0.747(9)   & \ \ 0.97 \ \ & 6 
\end{tabular}
}
\caption{Parameters of the $\chi^2$ fits of the transition lines in 
the $(\bar{y},\beta)$ [or $(u,\beta)$] plane for different values of 
$x$, extracted from simulations on the $L=6$ and $L=8$ lattices.}
\end{table}

\begin{figure}[t!]
\centerline{\includegraphics*[width=5.4in,angle=90]{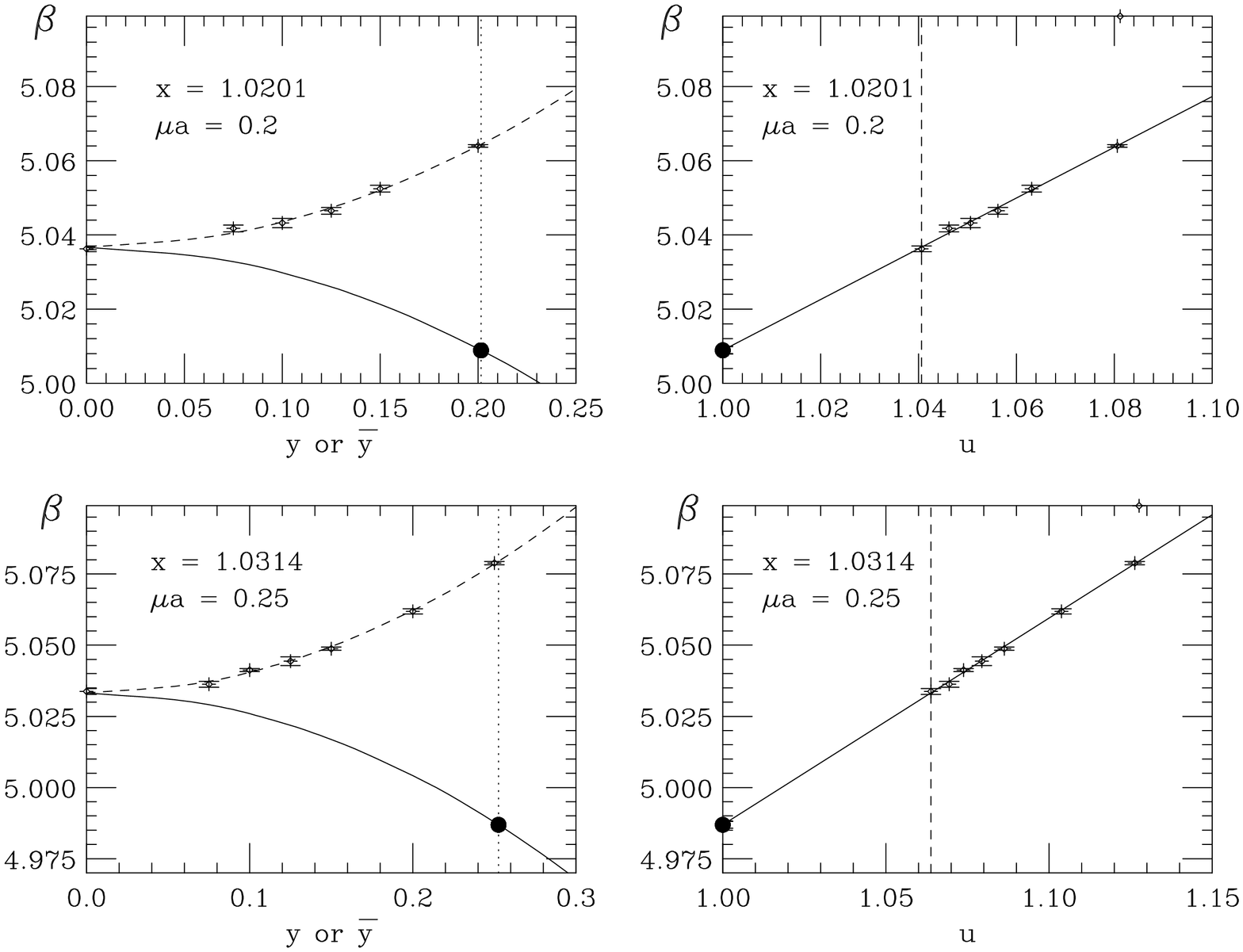}}
\caption{Phase diagrams at fixed $x$ in the $(\bar{y},\beta)$ and
$(y,\beta)$ planes (left panels) and in the $(u,\beta)$ plane
(right panels).}
\label{fig:phd_imag1}
\end{figure}

\begin{figure}[t!]
\centerline{\includegraphics*[width=2.7in,angle=90]{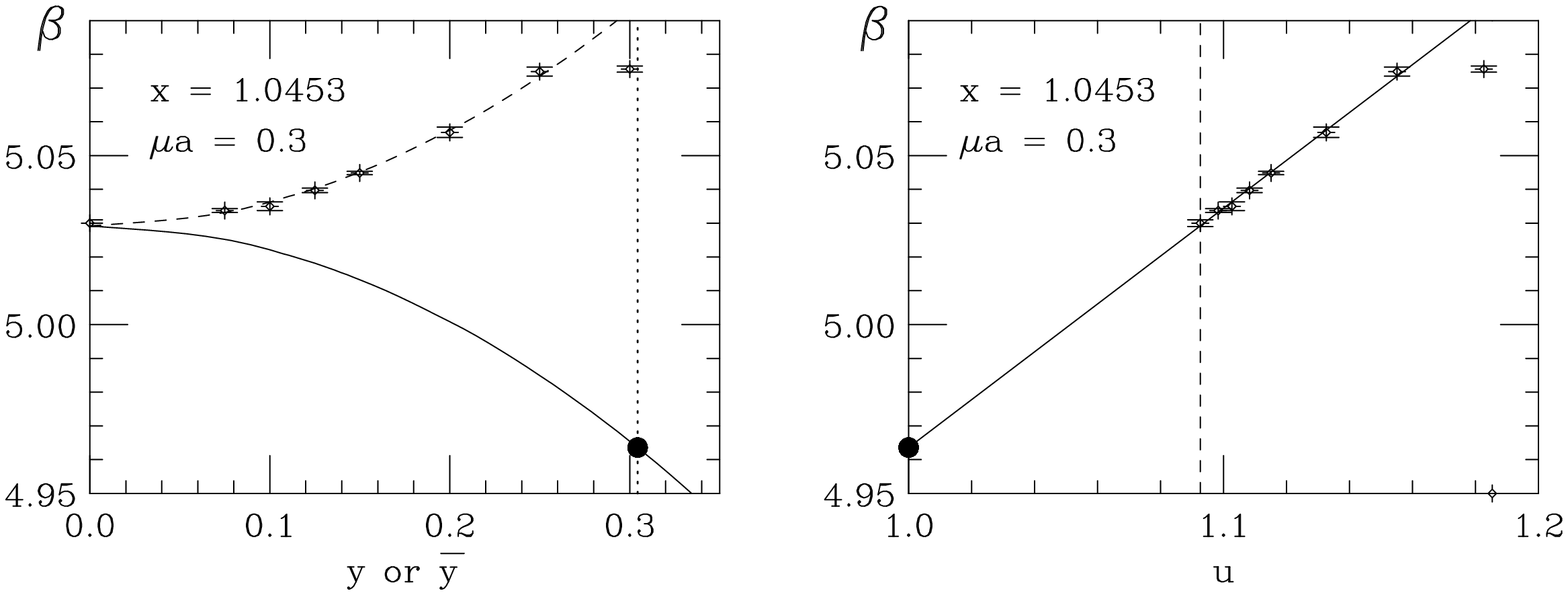}}
\centerline{\includegraphics*[width=5.4in,angle=90]{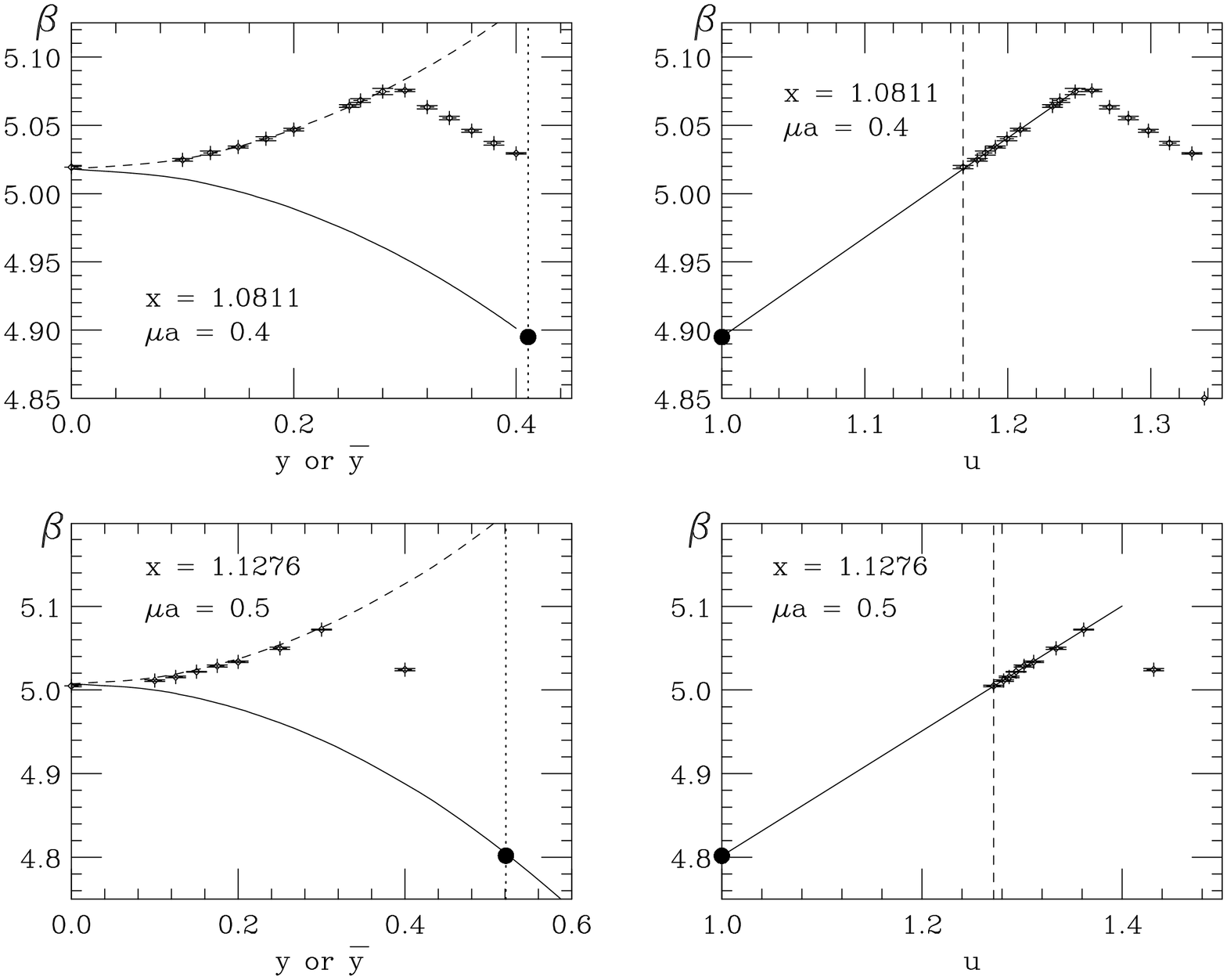}}
\caption{Phase diagrams at fixed $x$ in the $(\bar{y},\beta)$ and
$(y,\beta)$ planes (left panels) and in the $(u,\beta)$ plane
(right panels).}
\label{fig:phd_imag2}
\end{figure}

Table I collects the parameters of the best 
$\chi^2$ fits of the transition
lines given by equations (\ref{imag_line}) or (\ref{uline})
extracted from simulations on the $6^3\times 4$ and $8^3\times 4$.
The transition lines at fixed $x$ are displayed in Figures~\ref{fig:phd_imag1}
and~\ref{fig:phd_imag2}.
The left panels represent the $(\bar{y},\beta)$ plane,
for the open symbols and the dashed line, and the $(y,\beta)$ plane,
for the solid and dotted lines, and the filled circle.
The right panels represent the $(u,\beta)$ plane. 
In all plots the open symbols are the numerical
estimates of the phase transition points
obtained form the simulations on the $8^3\times 4$ lattice.
On the left panels, the dashed line is the best $\chi^2$ fit to a function
of the form (\ref{imag_line}) and the solid line is its analytical
continuation to $y=\mathrm{i} \bar{y}$. The dotted line is the physical
line, given by $y=y_\mathrm{ph}=\sqrt{x^2-1}$, and
the filled circle is the physical transition point.
On the right panels, the solid line is the fit of the transition points 
to a function of the form (\ref{uline}) and
the dashed line is the boundary of the region where numerical simulations
without sign problem can be performed (i.e., the line $u=x^2$).
The filled circle is the physical transition point obtained extrapolating
the transition line to the physical region, $u=1$.

In the transition lines displayed in Fig.~\ref{fig:phd_imag2}
the points at the largest values of $\bar{y}$ (or $u$) deviate from the 
smooth growing of the points at smaller $\bar{y}$ (or $u$), and they 
are not included
on the fits. They are a reflection of the $2\pi/3L_t$
periodicity that we would see
if we look at the phase diagram onto the constant $\beta$ plane 
instead of onto the plane 
of constant $x$ (see section 2).
The data of Fig. 6 (especially those for $x=1.0811$, for which many points
of the phase transition line were determined) strongly suggest that there 
is a cusp separating the $\beta$ increasing curve from the $\beta$ 
decreasing one. This means that the $2\pi/3L_t$ periodicity is realized
by the periodic replication of a nonperiodic analytic function,
producing a cusp at the points where a replica ends and a new replica starts. 
This cusp is a non-analyticity of the phase transition line which 
has the same origin as the
Roberge-Weiss transition \cite{rw} found in QCD at 
$\mu = \pm \mathrm{i}\pi/3L_t$
in the high temperature region \cite{alis}. Indeed, as it was conjectured in
\cite{ptp}, this singular behavior realized as a cusp is to be expected 
in a wide variety of
models characterized by a quantized charge coupled to a phase. In our case
the quantized charge is the baryonic charge and the phase
$\phi= \arctan (\bar{y}/x)$. Under this conditions, one 
must exclude the points to the right of the cusp from the fits since they 
belong to a different analytic function. Notice also that the presence of the 
nonanalytic cusp does not necessarily imply that the convergence radius of 
the Taylor series around $y=0$ ($u=1$) is limited by this singularity. 
Reference \cite{ptp} contains a simple illustrative example on that in what 
we called gaussian model.

The transition lines displayed in Figures~\ref{fig:phd_imag1}
and~\ref{fig:phd_imag2} are very smooth [they are essentially straight lines 
in the ($u, \beta$) plane until the cusp] and suggests a linear fit in 
the ($u, \beta$) plane. In fact the data reported in Table~I imply  
actually a very high confidence level for the fits and to 
add higher order corrections, as for instance a quartic term in equation 
(\ref{imag_line}), seems meaningless. In other words, if higher order 
corrections were relevant at real $y$, it would be very hard to measure them 
from the data produced at imaginary $y$. This makes really difficult any 
serious analysis of systematic errors possibly induced by the fit ansatz.

\begin{figure}[t!]
\centerline{\includegraphics*[width=3.5in,angle=90]{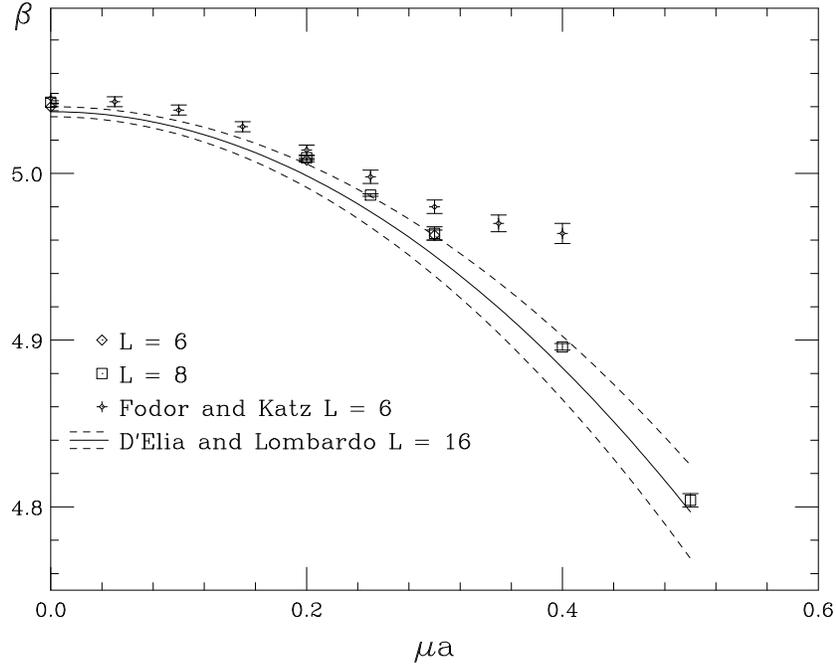}}
\caption{Phase diagram in the $(\mu a,\beta)$ plane. The solid line
is the analytical continuation of the imaginary chemical potential
pseudotransition line, and the dashed lines mark the error band. 
The authors of Ref.~\protect{\cite{mp}} give credit
to this analytical continuation up to $\mu a \approx 0.3$.}
\label{fig:phd_all}
\end{figure}

Figure~\ref{fig:phd_all} displays the phase diagram in the plane 
$(\mu a,\beta)$, together with the results obtained by 
Fodor and Katz with double reweighting \cite{fodor}, and
by D'Elia and Lombardo with imaginary chemical potential simulations
\cite{mp}. There is good agreement with the results of Fodor and 
Katz until $\mu a\approx 0.3$, especially if we take into account that
different methods to locate the pseudotransition coupling at finite
volume are used. 
The agreement with D'Elia and Lombardo is very satisfactory in the 
whole
range of $\mu a$ explored. The main difference, which is already seen at
$\mu a=0$, should be attributed to a volume effect, since these authors
used a $16^3\times 4$ lattice. D'Elia and Lombardo give credit to their
results for $\mu a\lesssim 0.3$, which is the interval where they found 
agreement with Fodor and Katz. We have seen that indeed the results 
(ours and theirs) are reliable
at least until $\mu a=0.5$, which is the maximum chemical potential
at which we performed simulations. The disagreement with Fodor and Katz 
for $\mu a\gtrsim 0.3$ is likely due to the poor overlap of the 
ensemble of reweighted
configurations with the ensemble of typical configurations at 
$\mu a\gtrsim 0.3$.

\begin{figure}[t!]
\centerline{\includegraphics*[width=3.5in,angle=90]{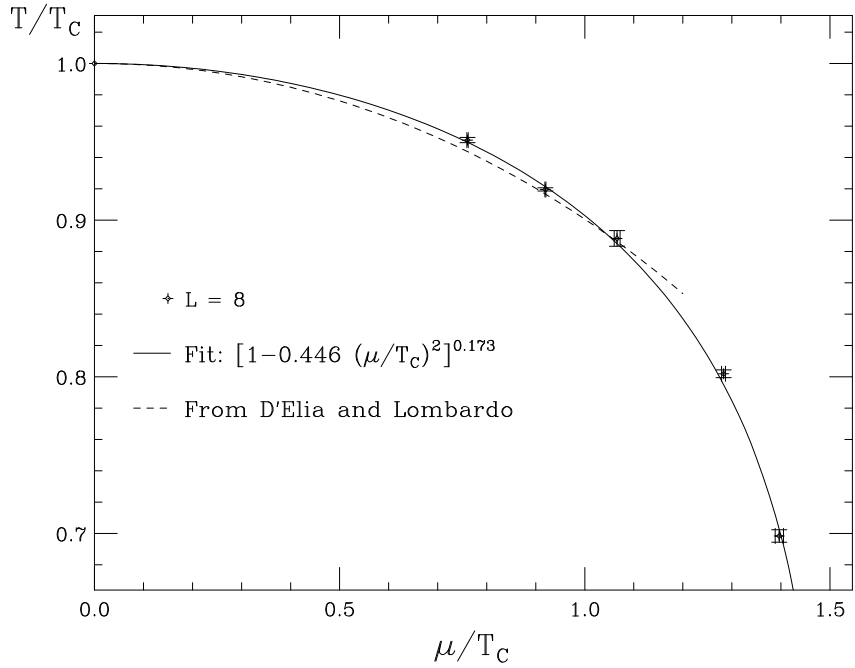}}
\caption{Phase diagram in physical units, with the zero density
transition temperature, $T_\mathrm{C}$, setting the scale.}
\label{fig:phd_phys}
\end{figure}

Figure~\ref{fig:phd_phys} shows the phase diagram in the plane $(\mu,T)$
in physical units, with the scale set by the transition temperature
at $\mu=0$, $T_\mathrm{C}$. The relative lattice spacings have been
determined by means of the two loop beta function. For comparison, 
we also plot the result of D'Elia and Lombardo. Our results
can be fit with a power function of the form

\begin{equation}
\frac{T}{T_\mathrm{C}}\;=\;\left[1-c \,(\mu/T_\mathrm{C})^2\right]^p\, .
\end{equation}

\noindent
The best $\chi^2$ fit gives $c\approx 0.446(9)$ and an exponent, 
$p\approx 0.173(7)$. One may be curious about the extrapolation of
this line to zero temperature. If this were done,
we would find a zero temperature transition at chemical potential
$\mu_\mathrm{C} \approx 1.5 T_\mathrm{C}$, which, in terms of the
nucleon mass \cite{espec}, $m_\mathrm{N}$, is 
$\mu_\mathrm{C} \approx m_\mathrm{N}/5$.

\section*{\normalsize 4 \ Conclusions}

Using a recently proposed method to determine the phase transition
lines at finite baryon density, we have obtained the phase transition 
line of QCD with four degenerate quark flavors from the zero density
high temperature transition, at $T_\mathrm{C}$, down to
$T\approx 0.7 T_\mathrm{C}$ and $\mu\approx 1.4 T_\mathrm{C}$. 
Our results are in reasonably good agreement with those of D'Elia and Lombardo,
obtained from simulations at imaginary chemical potential.
These authors give credit to their results for $T\gtrsim 0.9 T_\mathrm{C}$,
since this is the region where their results showed reasonably small 
statistical and systematic errors and in addition agree with those 
obtained by Fodor
and Katz with the double reweighting method. Our results agree reasonably
well with the central value reported by D'Elia and Lombardo in the whole 
region $T \gtrsim 0.7 T_\mathrm{C}$ and $\mu \lesssim 1.4 T_\mathrm{C}$ . 

We believe we can explore much lower temperatures with our method.
Since it is based on analytical continuation/extrapolation there are
uncontrolled systematic errors that grow in decreasing the temperature.
It is therefore very difficult to estimate the minimum temperature at
which our method will give a reliable prediction of the location of the phase 
transition point. In \cite{JHEP} we verified that in the three dimensional 
Gross-Neveu model at large $N$ this minimum temperature is amazingly low.

The study of four flavor QCD thermodynamics at lower temperatures is
very interesting, and we left it for future work. Since our method is
computationally very expensive, we decided to concentrate the present
effort in the more interesting cases of two and two plus one flavor
QCD.

\vspace{0.5truecm}

\noindent
\textit{Acknowledgments}

This work received financial support from CICyT (Spain),
project FPA2003-02948, from Ministerio de Ciencia y Tecnolog\'{\i}a 
(Spain), project BFM2003-08532-C03-01/FISI, and from
an INFN-CICyT collaboration. 
The authors thank the Consorzio Ricerca Gran Sasso that has provided
most of the computer resources needed for this work.
V.L. is a Ram\'on y Cajal fellow.

\end{document}